\documentclass[preprint,showpacs,preprintnumbers,amsmath,amssymb]{revtex4}

\usepackage{graphicx}% Include figure files
\usepackage{dcolumn}% Align table columns on decimal point
\usepackage{bm}% bold math

\begin{document}

\title{ Perturbative Calculation of Excluded Volume Effect \\
for Nuclear Matter in a Relativistic Mean-field Approximation }
\author{Bao-Xi SUN$^{1,2}$,
Xiao-Fu LU$^{2,3}$, En-Guang ZHAO$^{2,4,5}$}

\affiliation{${}^{1}$Institute of High Energy Physics, The Chinese Academy of
Sciences, P.O.Box 918(4), Beijing 100039, China}

\affiliation{${}^{2}$Institute of Theoretical Physics, The Chinese Academy of
Sciences, Beijing 100080, China}

\affiliation{${}^{3}$Department of Physics, Sichuan University, 
Chengdu  610064, China}

\affiliation{${}^{4}$Center of Theoretical Nuclear Physics, \\
National Laboratory of Heavy ion Accelerator,Lanzhou 730000,  China}

\affiliation{${}^{5}$Department of Physics, Tsinghua University,
Beijing 100084, China}

\begin{abstract}
Considering the finite volume of nucleons,
a Lagrangian density is given.
The first order self-energy of the nucleon
and the equation of
state of nuclear matter are calculated in the framework of relativistic
mean-field approximation. Our results indicate the effects of the
volume of nucleons are not negligible. \\
{\bf keywords:} excluded volume effect,  nuclear matter,
relativistic mean-field approximation,  nucleon-nucleon interaction.
\end{abstract}


\maketitle

\section{Introduction}
\par
Although the quantum field theory has succeeded in many areas of modern
physics, people have met some difficulties when the quantum field theory
is applied to  nuclear systems. Firstly, nuclear matter and all nucleus
are many-body systems, and their ground states are  states where
the Fermi sea is filled with interacting nucleons, not the "vacuum."
Secondly, the nucleons are extended objects, but they are
treated as "points" in the quantum field theory. These approximations
might cause some troubles in the research of nuclear physics.
Thus people have tried to exclude the whole volume ocuppied by nucleons
from their configuration space, then consider nucleons as "point-like"
particles moving in the mean field \cite{CR.86,HS.86,KT.89,
RG.91,ZMG.92,MZ.93,AS.95,TS.98}.
Since the nucleon volume is not relativistic invariant,
in ref\cite{Z.95,ZZ.95} a relativistic consistent volume effect of nucleons
is included in their calculation of the properties of nuclear matter.However,
all the previous methods of treating the finite volume effects are mainly
from the consideration of the geometry of the participating particles.

Now in this paper,
we incorporate the excluded volume effect into
the Lagrangian density of nuclear matter. Since the volume of nucleons
is about $10\%$ of the space in the saturation nuclear matter, we regard
the excluded volume effect as a perturbation and the first order
self-energy is calculated in the framework of relativistic mean-field
approximation. At last, a set of parameters is obtained so that the
saturation properties of nuclear matter can be reproduced well.

\section{The excluded volume effect in a Lagrangian density}
\par
To take into account of the excluded volume effect, the whole volume of nucleons
$v_{0}N$ is substracted from the volume of nuclear matter $V$ so that the
effective space available for  nucleon motion is $V-v_{0}N$,where $v_{0}$ is
the volume of a nucleon, $N$ is the total number of nucleons in the nuclear
matter.

\par
The nucleons in nuclear matter are constantly moving, constrainted by Pauli
principle. Thus the length of each nucleon decreases in the direction of its
movement by the Lorentz contraction and the effective space for nucleons 
is
\begin{equation}
V\left(1-v_{0}\rho_{s}^{\prime}(N)\right),
\end{equation}
where $\rho_{s}^{\prime}(N)$ is the scalar density of nucleons. The
box normalization factor becomes
\begin{equation}
\frac{1}{\sqrt{V\left(1-v_{0}\rho_{s}^{\prime}(N)\right)}},
\end{equation}
and the corresponding field is written as
\begin{equation}
\psi^{\prime}~=~\sqrt{1-v_{0}\rho_{s}^{\prime}(N)} \psi,
\end{equation}
where $\psi$ is the nucleon field in the effective space
$V\left(1-v_{0}\rho_{s}^\prime (N)\right)$. From now on,we use the quantities
with $"\prime"$ for the ones in the original space and the
quantities
 without $"\prime"$ for the quantities in the effective space.The
relation between the scalar densities of nucleons in the two spaces is
\begin{equation}
\rho_{s}^{\prime}(N)~=~\left(1-v_{0}\rho_{s}^{\prime}(N)\right)\rho_{s}(N).
\end{equation}
From the Eq.$(4)$,we obtain
\begin{equation}
\rho_{s}^{\prime}(N)~=~\frac{\rho_{s}(N)}{1+v_{0}\rho_{s}(N)}~=~
\rho_{s}(N)\left(1-v_{0}\rho_{s}(N)+v_{0}^{2}\rho_{s}^{2}(N)+...\right).
\end{equation}
If the $v_{0}^{2}$ terms and beyond are neglected, the above equation becomes
\begin{equation}
\rho_{s}^{\prime}(N)~=~\rho_{s}(N)\left(1-v_{0}\rho_{s}(N)\right).
\end{equation}

Since $\rho_{s}^{\prime}(N)$ and $\rho_{s}(N)$ are the scalar density
operators of nucleons in the original space and the effective space
respectively, namely $\rho_{s}^{\prime}(N)~=~\bar{\psi^{\prime}}\psi^{\prime}$,
~~~~$\rho_{s}(N)~=~\bar{\psi}\psi$, then we can obtain
\begin{equation}
\bar{\psi^{\prime}}\psi^{\prime}~=~\left(1-v_{0}
\bar{\psi}\psi\right)\bar{\psi}\psi.
\end{equation}
That is to say, when the volume of nucleons is excluded, the field operators
in the Lagrangian should be changed in the following way
\begin{equation}
\bar{\psi}O\psi\rightarrow\left(1-v_{0}\bar{\psi}\psi\right)\bar{\psi}
O\psi,
\end{equation}
where $O$ is an operator, such as 1, $\gamma^{\mu}$ etc.
The corresponding Lagrangian is written as
\begin{eqnarray}
{\cal L} & = &(1-v_{0}\bar{\psi}\psi)\bar{\psi}(i\gamma_{\mu}\partial^{\mu}
 -m-g_{\sigma}\sigma-g_{\omega}\gamma_{\mu}\omega^{\mu})\psi \nonumber \\
            &&+\frac{1}{2}\partial_{\mu}\sigma\partial^{\mu}\sigma-U(\sigma)
                -\frac{1}{4}\omega_{\mu\gamma}\omega^{\mu\gamma}+\frac{1}{2}
                m_{\omega}^{2}\omega_{\mu}\omega^{\mu},
\end{eqnarray}
where
\begin{equation}
U(\sigma)~=~\frac{1}{2}m_{\sigma}^{2}\sigma^{2}+\frac{1}{3}g_{2}\sigma^{3}
             +\frac{1}{4}g_{3}\sigma^{4}.
\end{equation}

The Lagrangian in the relativistic mean-field approximation is
\begin{eqnarray}
{\cal L}_{RMF}&=&(1-v_{0}\bar{\psi}\psi)\bar{\psi}(i\gamma_{\mu}\partial^{\mu}
 -m-g_{\sigma}\sigma_{0}-g_{\omega}\gamma^{0}\omega_{0})\psi \nonumber \\
&&-\frac{1}{2}m_{\sigma}^{2}\sigma_{0}^{2}-\frac{1}{3}g_{2}
\sigma_{0}^{3} -\frac{1}{4}g_{3}\sigma_{0}^{4}
+\frac{1}{2}m_{\omega}^{2}\omega_{0}^{2},
\end{eqnarray}
where $\sigma_0$ and $\omega_0$ are the expectation values of meson fields
in the ground states of nuclear matter.
 Since the radius of a nucleon in nuclear matter is usually taken as 0.63fm,
the whole volume
of nucleons is only about $10\%$ of the volume of the saturated nuclear
matter so that the effect of nucleon volume can be treated as a perturbation.
The Lagrangian density ${\cal L}_{RMF}$ can be divided into
nonperturbative and
perturbative parts, e.g.
\begin{equation} 
{\cal L}_{RMF}~=~{\cal L}_{0}~+~{\cal L}_{I}
\end{equation} 
with
\begin{eqnarray}
{\cal L}_{0}&=&\bar{\psi}(i\gamma_{\mu}\partial^{\mu}
 -m-g_{\sigma}\sigma_{0}-g_{\omega}\gamma^{0}\omega_{0})\psi \nonumber \\
&&-\frac{1}{2}m_{\sigma}^{2}\sigma_{0}^{2}-\frac{1}{3}g_{2}
\sigma_{0}^{3} -\frac{1}{4}g_{3}\sigma_{0}^{4}
+\frac{1}{2}m_{\omega}^{2}\omega_{0}^{2},
\end{eqnarray}
and
\begin{equation}
{\cal L}_{I}=-v_{0}\bar{\psi}\psi\bar{\psi}(i\gamma_{\mu}\partial^{\mu}
 -m-g_{\sigma}\sigma_{0}-g_{\omega}\gamma^{0}\omega_{0})\psi
\end{equation}
in the interaction picture
defined by the nonperturbative Lagrangian density ${\cal L}_{0}$.
The equation of motion for the nucleon field  can be easily obtained from
${\cal L}_{0}$:
\begin{equation}
\left(i\gamma_{\mu}\partial^{\mu}-m-g_{\sigma}\sigma-g_{\omega}\gamma_{\mu}
\omega^{\mu}\right)\psi~=~0.
\end{equation}

The perturbative
Hamilton in the interaction picture is
\begin{equation}
{\cal H}_{I}
~=~-v_{0}\bar{\psi}\psi\left(\bar{\psi}\left(-i\vec{\gamma}\cdot\vec{\nabla}
+m^{\ast}+g_{\omega}\gamma^{0}\omega_{0}\right)\psi\right),
\end{equation}
where $m^{\ast}~=~m+g_{\sigma}\sigma_{0}$.
The first order of the evolution operator $U(t_2,t_1)$ is
\begin{equation}
U(t_2,t_1)~=~1-i\int\limits_{t_{1}}^{t_{2}}dt\int d^{3}x {\cal H}_{I}
                 (\vec{x},t).
\end{equation}
When the effect of nucleon volume is considered,
the corresponding Green function is
\begin{eqnarray} 
G~&=&~\langle 
0\left|T\{\psi(x_{2})\bar{\psi}(x_{1})U(+\infty,-\infty)\}\right|0     
\rangle  \nonumber  \\ ~&=&~\langle 
0\left|T\{\psi(x_{2})\bar{\psi}(x_{1})\}\right|0\rangle  \\ 
&&-v_{0} \int\limits_{-\infty}^{+\infty}dt\int d^{3}x      \langle 
0\left|T\{\psi(x_{2})\bar{\psi}(x_{1})\bar{\psi}(x)\gamma_{0}      
\psi(x)\bar{\psi}(x) \left(-\vec{\gamma}\cdot\vec{\nabla}-im^{\ast}-      
ig_{\omega}\gamma^{0}\omega_{0}\right)\psi(x)\}\right|0\rangle. \nonumber 
\end{eqnarray} 
After some straightforward derivation the first order self-energy of nucleon
can be expressed as
\begin{equation}
\Sigma~=~I\Delta m+\gamma_{0}\Sigma_{0}-\vec{\gamma}\cdot\vec{\Sigma},
\end{equation}
where
\begin{equation} 
\Delta m~=~v_{0}\int\frac{d^{3}k}{(2\pi)^{3}}\left(\frac{3}{2}E^{\ast}(k)           
+g_{\omega}\omega_{0}+\frac{3}{2}\frac{m^{\ast^{2}}}{E^{\ast}(k)}\right), 
\end{equation} 
 \begin{equation}
\Sigma_{0}~=~-v_{0}\int\frac{d^{3}k}{(2\pi)^{3}}\left(m^{\ast}-
            \frac{g_{\omega}\omega_{0}}{E^{\ast}(k)} m^{\ast}\right),
\end{equation}
\begin{equation}
\vec{\Sigma}~=~-v_{0}\int\frac{d^{3}k}{(2\pi)^{3}}\left(\frac{3}{2}
             \frac{m^{\ast}}{E^{\ast}(k)}\right)\vec{k_{1}}
\end{equation}
with $E^{\ast}(k)~=~\sqrt{\vec{k^{2}}+m^{\ast^{2}}}$.
In the first order approximation, the Green function reads as
\begin{equation}
G(k_{1})~=~G^{0}(k_{1})-iG^{0}(k_{1})\Sigma G^{0}(k_{1})
\end{equation}
with
\begin{equation}
G^{0}(k_{1})~=~\left(\frac{i}{\rlap{/}{k_{1}}-m^{\ast}+i\varepsilon}
               \right)_{\beta\alpha}.
\end{equation}
This Green function can be rewritten as following
\begin{eqnarray}
G(k_{1})~&=&~\left(\frac{i}{\rlap{/}{k_{1}}-m^{\ast}-\Sigma+i\varepsilon}
               \right)  \nonumber  \\
&=&~\left(\frac{i}{ \gamma_0
(\varepsilon^{\ast}(k_{1})-\Sigma_{0})-\vec{\gamma}\cdot(\vec{k_{1}}-\vec{\Sigma})-
(m^{\ast}+\Delta m) +i\varepsilon}
               \right)  \nonumber \\
&=&~\left(\frac{i}{ \gamma_0
(\varepsilon^{\ast}(k_{1})-\Sigma_{0})-\vec{\gamma}\cdot\vec{k_{1}}X-
(m^{\ast}+\Delta m) +i\varepsilon}
               \right).
\end{eqnarray}
namely,
\begin{equation}
G(k_{1})~=~\left(\frac{i X^{-1}}{ \gamma_0
(\frac{\varepsilon^{\ast}(k_{1})-\Sigma_{0}}{X})-\vec{\gamma}\cdot\vec{k_{1}}-
(\frac{m^{\ast}+\Delta m}{X}) +i\varepsilon}\right),
\end{equation}
in which
\begin{equation}
X=1+\frac{3}{8}v_0 \rho_s(N).
\end{equation}
Now we can see that the perturbation interaction causes the corrections
of mass,energy and wave function of the nucleon.
which are expressed by $\Delta m$, $\Sigma_0$
and $X^{-1}$.
The effective mass $m^{\ast}$ and energy of a nucleon
$\varepsilon^{\ast} (k_{1})$ should be changed as
\begin{equation}
m^{\ast}\rightarrow \frac{m^{\ast}+\Delta m}{X},
\end{equation}
\begin{equation}
\varepsilon^{\ast}(k_{1})\rightarrow\frac{\varepsilon^{\ast}(k_{1})-\Sigma_{0}}{X}.
\end{equation}
The scalar density $\rho_{s}(N)$ and vector density $\rho_{v}(N)$
of nucleons are
\begin{equation}
\rho_{v}(N)~=~\frac{4}{(2\pi)^{3}}\int\limits_{0}^{k_{F}}d^{3}k,
\end{equation}
\begin{equation}
\rho_{s}(N)~=~\frac{4}{(2\pi)^{3}}\int\limits_{0}^{k_{F}}d^{3}k
              \frac{m^{\ast}}{\sqrt{(\vec{k})^{2}+
              m^{\ast^{2}}}}.
\end{equation}

We must calculate  the effective mass $m^{\ast}$ of nucleons
self-consistently:
\begin{eqnarray}
m^{\ast}~&=&~\frac{1}{X}\left( m+\Delta m \right) \nonumber \\
&&-\frac{1}{X}C_{s}^{2}\left(\frac{4}{(2\pi)^{3}}
\int\limits_{0}^{k_{F}} d^{3}k\frac{Xm^{\ast}-\Delta
m}{(\vec{k}^{2}+(Xm^{\ast}-\Delta m)^{2})
^{\frac{1}{2}}}\right) \nonumber \\
&&-\frac{1}{X}C_{s}^{2}\left(B(Xm^{\ast}-\Delta
m-m)^{2}+C(Xm^{\ast}-
           \Delta m-m)^{3})\right),
\end{eqnarray}
where
$~C_{s}~=~g_\sigma/m_\sigma~,$ $~B=g_2/g_\sigma^3~,$ $~C=g_3/g_\sigma^4~$ 
and
 \begin{equation}
 k_{F}~=~\left(\frac{3\pi^{2}}{2}\rho_{v}(N)\right)^{\frac{1}{3}}.
 \end{equation}
From the equations
\begin{equation}
 m^2_\sigma \sigma_0 + g_2 \sigma_{0}^{2} + g_3 \sigma_{0}^3
 =-g_\sigma\rho_s(N)
\end{equation}
and
\begin{equation}
m^2_\omega \omega_0 =g_\omega\rho_v(N),
\end{equation}
the values of $\sigma_0$ and $\omega_0$ are calculated. At last
we acquire the
energy density of symmetric nuclear matter 
\begin{equation}
\varepsilon~=~\varepsilon_N ~+~\frac{1}{2} m_{\sigma}^2 \sigma_0^2
~+~\frac{1}{3}g_{3}\sigma_0^3~+~\frac{1}{4}g_{4}\sigma_0^4 -
\frac{1}{2} m^2_\omega \omega^2_{0},
\end{equation}
where
\begin{eqnarray}
\varepsilon_N &=& \frac{4}{\left( 2\pi \right)^3} \int^{k_{F}}_{0}
d\vec{k}
 \left(\vec{k}^2 + {m^{\ast}}^2 \right)^{\frac{1}{2}} \nonumber \\
&&+\frac{2}{3\pi^2} {k_{F}}^3 \left(g_\omega
\omega_0~+~\frac{1}{4}~v_0~ \left(m^\ast \rho_v (N)~-~g_\omega
\omega_0 \rho_S (N) \right) \right).
\end{eqnarray}

\section{The equation of state for nuclear matter}
In our calculation the radius of the nucleon in the nuclear matter is
taken as $0.63fm$ and the
four parameters $g_{\sigma}$, $g_{\omega}$, $g_{2}$, $g_{3}$ in this model
are fixed by fitting the saturation properties of normal nuclear matter.
The parameters in this
model(labeled by ITP)are listed in Table 1. As a comparison,the parameter set
NL3, which is often used in the RMF calculation with all the nucleons taken as
point-like particles\cite{P.97}, and QRZ set, which is obtained in
ref\cite{Sun.01} by
 including
 the finite volume effect from the geomitry
consideration, are also
 given
 in the table 1.
Comparing with the NL3 set, the coupling constant to scalar mesons $g_\sigma$
of ITP becomes lager .Since the excluded volume effect actually
supplies a repulsive force,so this trend is reasonable.
For the coupling constant to vector mesons $g_\omega$ which is associated with
the repulsive interaction between nucleons, its ITP value is smaller
than the one of NL3. This is consistent with the result of QRZ set.

\par
With different set of parameter, the average energy per nucleon as a function
of the number density for nuclear matter is shown in Fig.1.
Comparing these results, we can see that the average energy per nucleon of
our calculation
increases more quickly than those of NL3\cite{P.97} and
QRZ\cite{Sun.01} as $\rho_S(N)>2\rho_0$ although the values of
compression modulus are almost the same at the saturation
density. It manifests that the influence of the excluded volume
effect of nucleons increases at higher density. The results at
even higher densities, are not given in the figure since we are not sure
if the purturbative model used in this paper
is still suitable to calculate the equation of state at those densities .

\par
The effective nucleon mass  $M^{\ast}_N$
as a function of the
nucleon density $\rho_N$ for nuclear matter with different set of
parameters is displayed in Fig.2. 
It decreases as the number density of nucleons increases. However,
a little larger effective nucleon mass is obtained in our model
than the result of the relativistic mean-field approximation where
the nucleon is treated as "point-like" particles with NL3 parameters
but the change is not so significant as indicated by QRZ curve.
The differences between the three models become larger as the density
increases.This implies that at higher densities we have not only to
take into account of the finite volume effect of nucleons but also
have to pay attention to choose a "better" model to deal with it.

\section{Summary}
In summary, we give a Lagrangian density including the effect of
volume of nucleons,
then the first order self-energy of the nucleon is derived 
, and the equation of
state of nuclear matter is calculated in the framework of relativistic
mean-field approximation. Our results indicate that the finite
volume can cause a considerable influence on some properties of the
nuclear matter.

\begin{acknowledgments}
The authors are grateful for the discussion with Z.X.Chang and F.Wang
and the supports from the National Natural Science Foundation of China
, the Major State Basic Research Development Programme under Contract
No.G2000-0774 and the CAS Knowledge Innovation Project No. KJCX2-N11.
\end{acknowledgments}

\begin{table} [h]
\caption{The parameters in the calculation of relativistic
mean-field approximation} \vspace*{0.5cm}
\begin{center}
\begin{tabular}{|c|c|c|c|}\hline
                   & ITP$\left(This~model\right)$&NL3\cite{P.97}&QRZ\cite{Sun.01}\\ \hline
$M$~$(MeV)$        &  939      &  939      & 939      \\ \hline
$m_\sigma$~$(MeV)$ &  550      &  508.194  & 532      \\ \hline
$m_\omega$~$(MeV)$ &  783      &  782.501  & 780      \\ \hline
$g_\sigma$         &  12.30    &  10.217   & 7.676    \\ \hline
$g_\omega$         &   6.45    &  12.868   & 7.036    \\ \hline
$g_2$~$(fm^{-1})$  & -17.30    &  -10.431  & 28.140   \\ \hline
$g_3$              &  28.00    &   -28.885 &  0.012   \\ \hline
$r_c$~$(fm)$       &  0.63     &   0.00    & 0.62     \\ \hline
Nuclear matter properties \\ \hline $\rho_0$~$(fm^{-3})$  &  0.145
&  0.148  & 0.147     \\ \hline $E/A$~$(MeV)$         & 16.727   &
16.299 & 15.687     \\ \hline $K$~$(MeV)$           & 288.65   &
271.76 & 241     \\ \hline $m^{\ast}/m$          & 0.695    &
0.60   & 0.83     \\ \hline
\end{tabular}
\end{center}
\end{table}

\newpage

\leftline{\Large {\bf Figure Captions}} 
\parindent = 2 true cm 
\parskip 1 cm 
\begin{description}

\item[Fig. 1]  Average energy per nucleon $E/A$ as a function of the 
nucleon density $\rho_N$ for nuclear matter with different  
parameters, 
(ITP) for perturbative calculation of the excluded volume effect in this 
model, 
(NL3) for the relativistic mean-field approximation with NL3 parameter
\cite{P.97}, 
(QRZ) for the results in ref.\cite{Sun.01}. \\   
 
\item[Fig. 2]  Effective mass of nucleons $M^{\ast}_N$ 
as a function of the 
nucleon density $\rho_N$ for nuclear matter with different  
parameters, The meanings of ITP, NL3 and QRZ 
are same as those of Fig.1. \\ 
 
\par 
 
\end{description} 

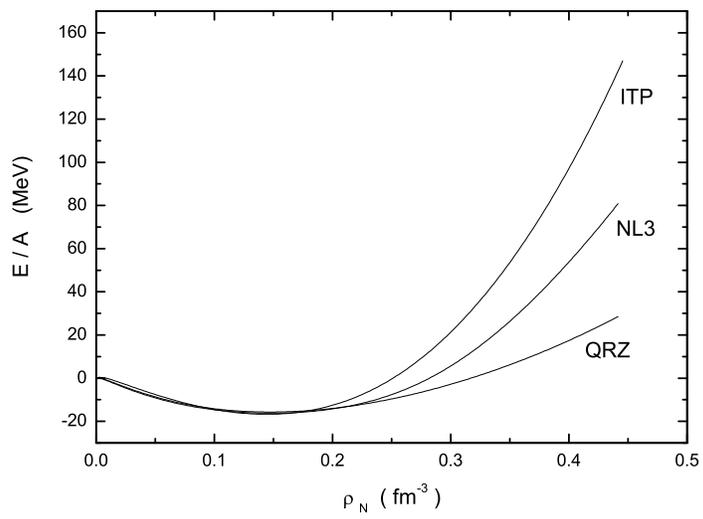
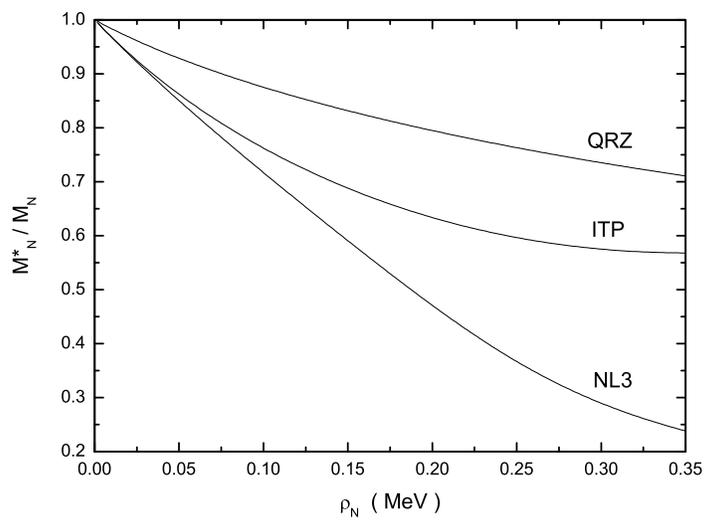
\end{document}